\documentclass[final,3p,times]{elsarticle}
\usepackage{ecrc}
\usepackage{url}
\usepackage{bm}
\volume{15}
\firstpage{92}
\journalname{Physics Procedia}
\runauth{M. Weigel and T. Yavors'kii}
\jid{phpro}
\jnltitlelogo{Physics Procedia}
\usepackage{amssymb}
\biboptions{square,comma,numbers,sort&compress}

\usepackage[figuresright]{rotating}

\begin{document}

\begin{frontmatter}



\dochead{}

\title{GPU accelerated Monte Carlo simulations of lattice spin models}


\author{M. Weigel}
\author{T. Yavors'kii}

\address{Institut f\"ur Physik, Johannes Gutenberg-Universit\"at Mainz,
  Staudinger Weg 7, D-55099 Mainz, Germany}

\begin{abstract}
  We consider Monte Carlo simulations of classical spin models of statistical
  mechanics using the massively parallel architecture provided by graphics processing
  units (GPUs). We discuss simulations of models with discrete and continuous
  variables, and using an array of algorithms ranging from single-spin flip Metropolis
  updates over cluster algorithms to multicanonical and Wang-Landau techniques to
  judge the scope and limitations of GPU accelerated computation in this field. For
  most simulations discussed, we find significant speed-ups by two to three orders of
  magnitude as compared to single-threaded CPU implementations.
\end{abstract}


\end{frontmatter}

% \linenumbers

\section{Introduction}

Due to their rich content in phase transitions and critical phenomena and the wealth
of applications in condensed-matter physics and beyond, classical spin systems have
been studied numerically in an enormous body of published research to date
\cite{binder:book2}. The ever increasing demand in computational power for computer
simulations of spin models has prompted researchers interested in these problems to
regularly use the cutting-edge computer technology of their time. In the past and
present this has included the design and implementation of special-purpose computers
for spin model simulations such as the cluster processor for studying critical
phenomena in the ferromagnetic Ising model \cite{bloete:99a} or, more recently, the
Janus machine for simulations of spin-glass models \cite{belleti:09}. The
construction of such machines entails an effort in financial and human resources
significantly exceeding that of programming a regular computer or compute
cluster. With the advent of general purpose computing on graphics processing units
(GPGPU) \cite{owens:08} a high-performance computing architecture is gaining more and
more widespread use in the scientific community \cite{hwu:11} that is special-purpose
in its original objective of displaying graphics and certainly less general in its
applicability and flexibility than current CPUs but, nevertheless, significantly more
easily accessible than the special-purpose machines mentioned above.

The possibility of harvesting the nominally vast computational power provided by
current-generation GPUs rests on the degree to which a number of challenges can be
met: (a) the efficient programming of such devices without the need of using (too)
low-level techniques, (b) the suitability of the problem and the algorithms employed
for calculations in a massively parallel setup, and (c) the degree to which
peculiarities of such devices, such as special hierarchies of cache memories,
difficulties with thread scheduling, or reduced precision in floating-point
arithmetics, can be catered for in the implementations \cite{kirk:10}. While a device
independent programming interface has recently become available with the OpenCL
framework \cite{opencl}, we here use the NVIDIA CUDA toolkit \cite{cuda} due to its
currently more mature status. It is the purpose of this contribution to summarize our
experience with the suitability of a number of popular algorithms for the simulation
of spin models for the massively parallel environment provided by current GPUs. In
the course of describing the results for various models and algorithms, we will
discuss the question in how far architectural limitations and peculiarities represent
significant hurdles for the efficient use of GPUs.

\section{Metropolis algorithm}

We studied classical O($n$) spin models with Hamiltonian
\begin{equation}
  \label{eq:hamiltonian}
  {\cal H} = -\sum_{\langle ij\rangle} J_{ij} \bm{s}_i\cdot \bm{s}_j,
\end{equation}
where $n=1$ corresponds to the discrete Ising model and $n=2$ and $n=3$ describe the
continuous {\em XY\/} and Heisenberg models, respectively. The spins are located on
square ($d=2)$ or simple cubic ($d=3$) lattices and interact with nearest neighbors
only. We first considered simulations using only single spin-flip moves accepted
according to the Metropolis criterion \cite{metropolis:53a},
\begin{equation}
  \label{eq:metropolis}
  p_\mathrm{acc}(s_i \mapsto s_i') = \min\left[1,\,e^{-\beta\Delta E}\right].
\end{equation}
We also considered updates according to the heat-bath prescription
\cite{binder:book2} which lead to only slightly (and locally) modified implementations
and very similar speed-ups on GPUs as compared to CPU implementations and are,
therefore, not discussed in detail here.

\begin{figure}[tb]
  \centering
  \raisebox{0.775cm}{\includegraphics[width=0.4\textwidth]{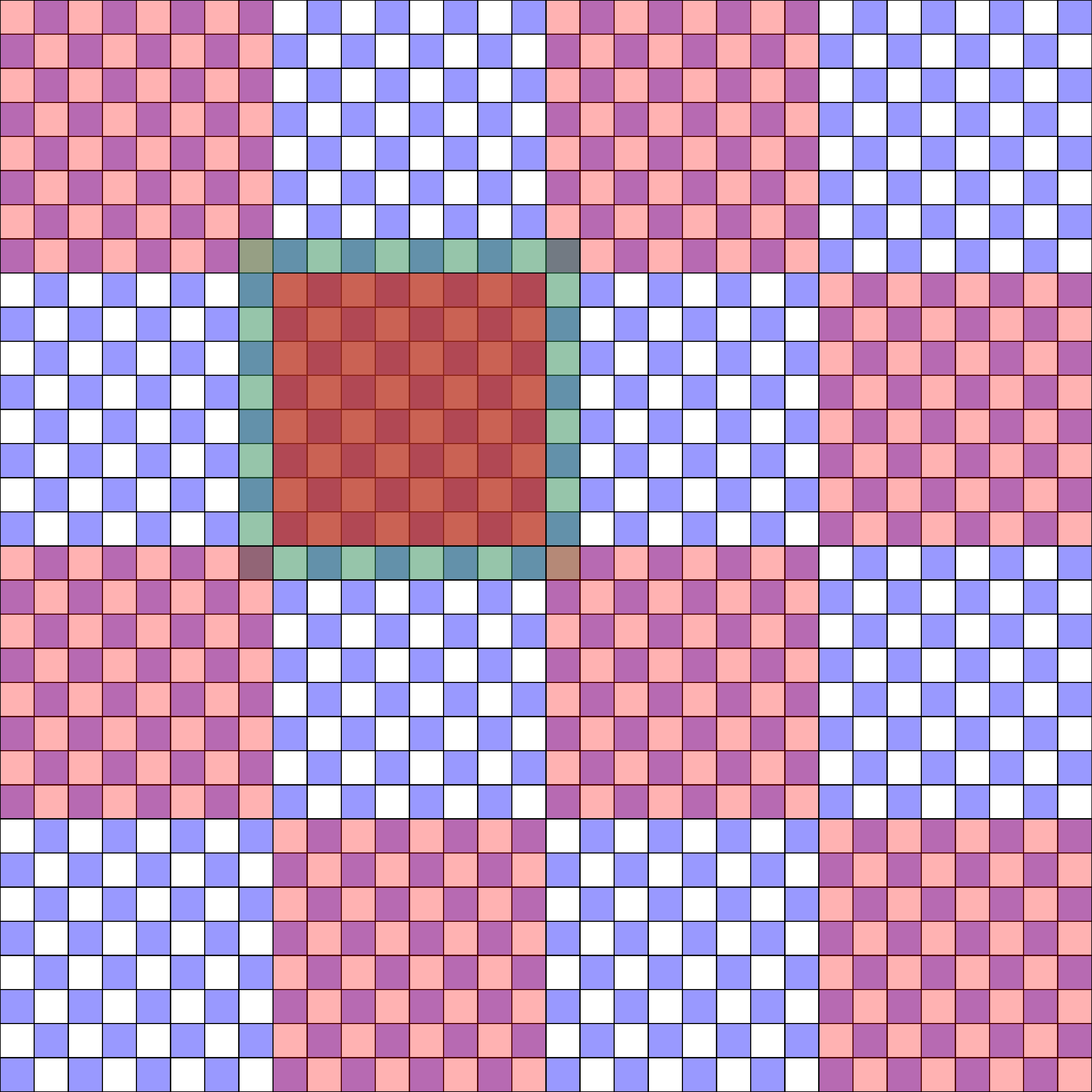}}
  \hspace*{0.75cm}
  \includegraphics[keepaspectratio=true,scale=0.65,trim=45 48 75 78]{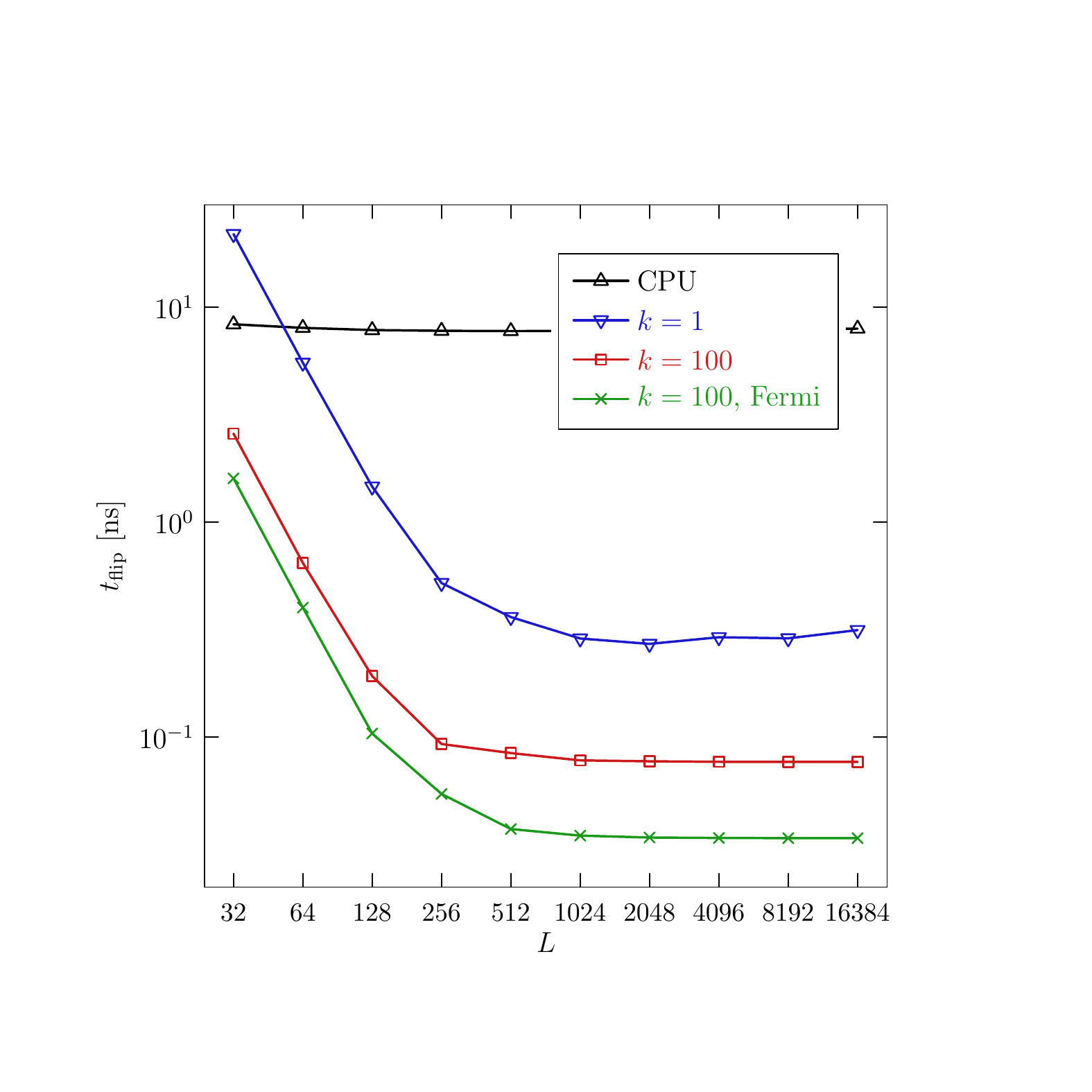}
  \caption{Left: double checkerboard decomposition of a square lattice of edge length $L =
    32$ for performing a single spin-flip Metropolis simulation of spin models on
    GPU. Right: Time per spin-flip in the 2D Ising model on CPU and on GPU with different
    choices of $k$. GPU data are for the Tesla C1060 device apart from the lowest
    curve which is for a GTX 480 card.}
  \label{fig:checker}
\end{figure}

For an implementation on GPU, one needs to allow for the parallel update of a large
number of spins. This is most straightforwardly accomplished for the case of
nearest-neighbor interactions by making use of a checkerboard decomposition of the
lattice \cite{heermann:90}. To make efficient use of {\em shared memory\/} that can
be accessed concurrently with very low latencies from all threads on a
multiprocessor, we use a two-level hierarchical decomposition, cf.\ the left panel of
Fig.~\ref{fig:checker} \cite{weigel:10a,weigel:10c}. All spins of one of the big
tiles plus some boundary layer are collaboratively loaded into shared memory and
subsequently updated by the individual threads of a thread block
\cite{kirk:10}. Inside of each tile, the checkerboard arrangement allows for all
spins of one sub-lattice to be updated concurrently before a synchronization barrier
occurs and the second sub-lattice is treated analogously. To amortize the effort of
loading tiles into shared memory, a number $k$ of updates of all spins of all big
tiles of one color is performed before updating the second sub-lattice. Depending on
the size of the tiles, this slows down the decorrelation of spin configurations. This
effect, however, is more than counter-balanced by the performance increase, even
close to criticality \cite{weigel:10c}. For good performance, a number of additional
tricks are employed, including a pre-tabulation of the Boltzmann factors in
Eq.~(\ref{eq:metropolis}) while storing this table as a texture \cite{kirk:10}, and
generation of random numbers even if they are not required to reduce thread
divergence. The simulation code can be downloaded at the authors' web site
\cite{weigel:gpu}. For the comparisons discussed here, an array of simple $32$-bit
linear congruential pseudo-random number generators (one per thread) is
used. Although these are known to have rather poor properties, for the purpose at
hand they appear to be sufficient even for high-precision results. The implementation
of more generally appropriate generators is discussed in Ref.~\cite{weigel:10c}.

For the benchmarks, we compared the performance of the outlined GPU implementation on
a Tesla C1060 as well as a more recent GTX 480 of the Fermi architecture series with
the results of an optimized, single-threaded CPU code running on an Intel Core 2 Quad
Q9650 at 3.0 GHz. For the Ising ferromagnet, $n=1$ and $J_{ij} = 1$ in
Eq.~(\ref{eq:hamiltonian}), we find a tile size of $16\times 16$ spins to be optimal
for sufficiently large systems \cite{weigel:10c}. The maximum performance reached on
the GTX 480 is around $0.03$ ns per spin flip (using $k=100$), which is $235$ times
faster than the CPU implementation, cf.\ the data collected in
Tab.~\ref{tab:performance} and the right panel of Fig.~\ref{fig:checker}. The Tesla
C1060, on the other hand, roughly performs at half of this speed. This speed-up,
however, is only reached for sufficiently large system sizes that allow to fully load
the $240$ and $480$ cores of the C1060 and the GTX 480, respectively. Very similar
relative performance is observed for the Ising model in three dimensions. For models
with continuous spins, exemplified by the 2D Heisenberg model with $n=3$ and $J_{ij}
= 1$ in Eq.~(\ref{eq:hamiltonian}), issues of floating-point performance and
precision become important. We find that a mixed-precision calculation, where the
spins are represented in single precision and only aggregate quantities such as the
total energy are calculated in double precision (see ``Metropolis single'' in
Tab.~\ref{tab:performance}) yield high performance without problems with
precision. If we employ the hardware optimized implementations of the special
functions (trigonometric, exponential, logarithmic etc.) provided in the CUDA
framework total speed-ups beyond $1000$ can be achieved as compared to CPU codes (see
``Metropolis fast math'' in Tab.~\ref{tab:performance}).

Simulations of systems with quenched disorder allow for trivial parallelization over
disorder realizations on top of the domain decomposition outlined above. For the
Edwards-Anderson Ising spin glass with couplings $J_{ij}\in\{-J,J\}$ drawn from a
bimodal distribution, we again find speed-ups of around $100$ and $200$ for the Tesla
C1060 and the GTX 480, respectively. Further improvements can be achieved on using
$64$-bit multi-spin coding which allows for spin-flip times down to $2$ pico-seconds
on the GTX 480, cf.\ Tab.~\ref{tab:performance}.

\begin{figure}[tb]
  \centering
  \includegraphics[keepaspectratio=true,scale=0.65,trim=45 48 75 78]{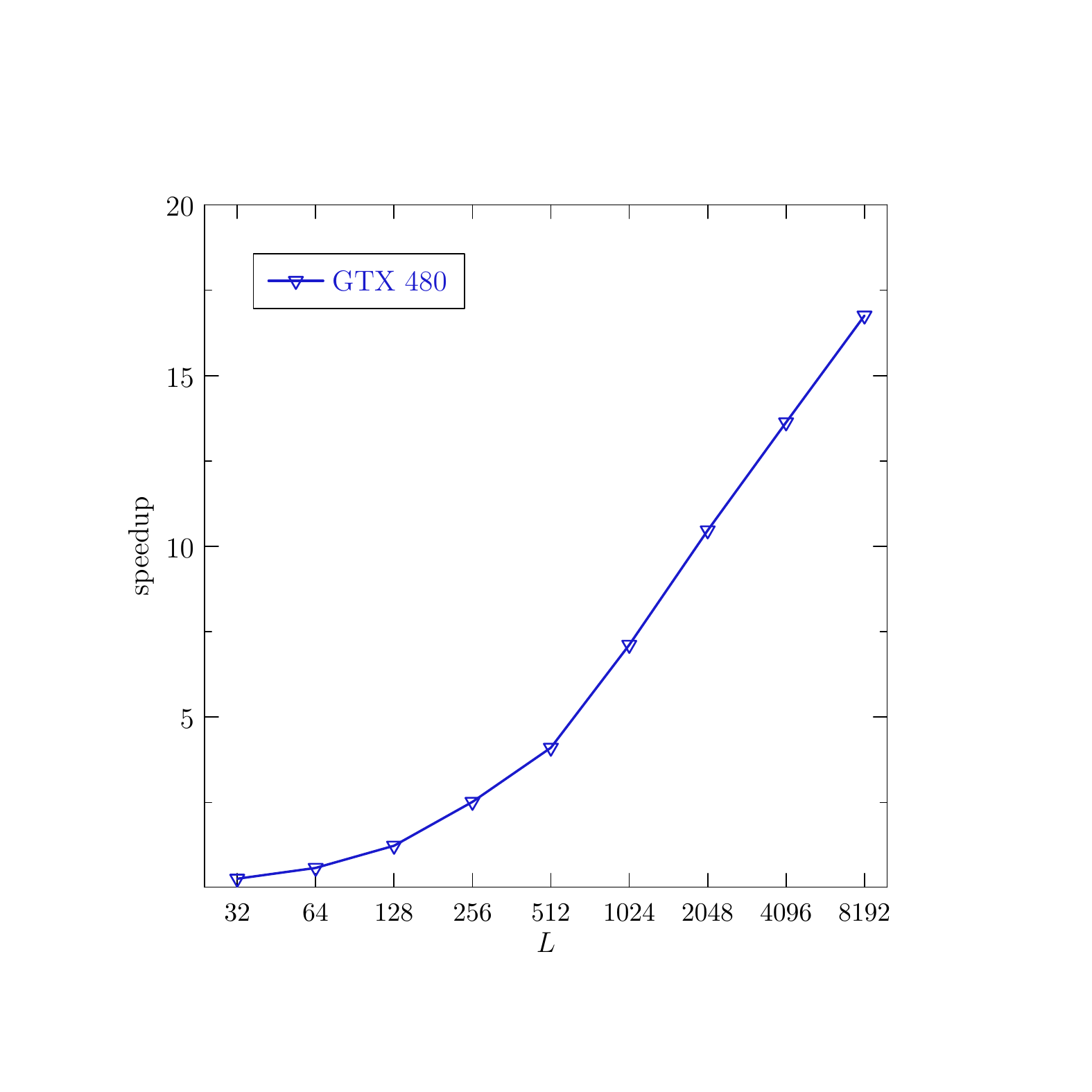}
  \hspace*{0.5cm}
  \includegraphics[keepaspectratio=true,scale=0.65,trim=45 48 75 78]{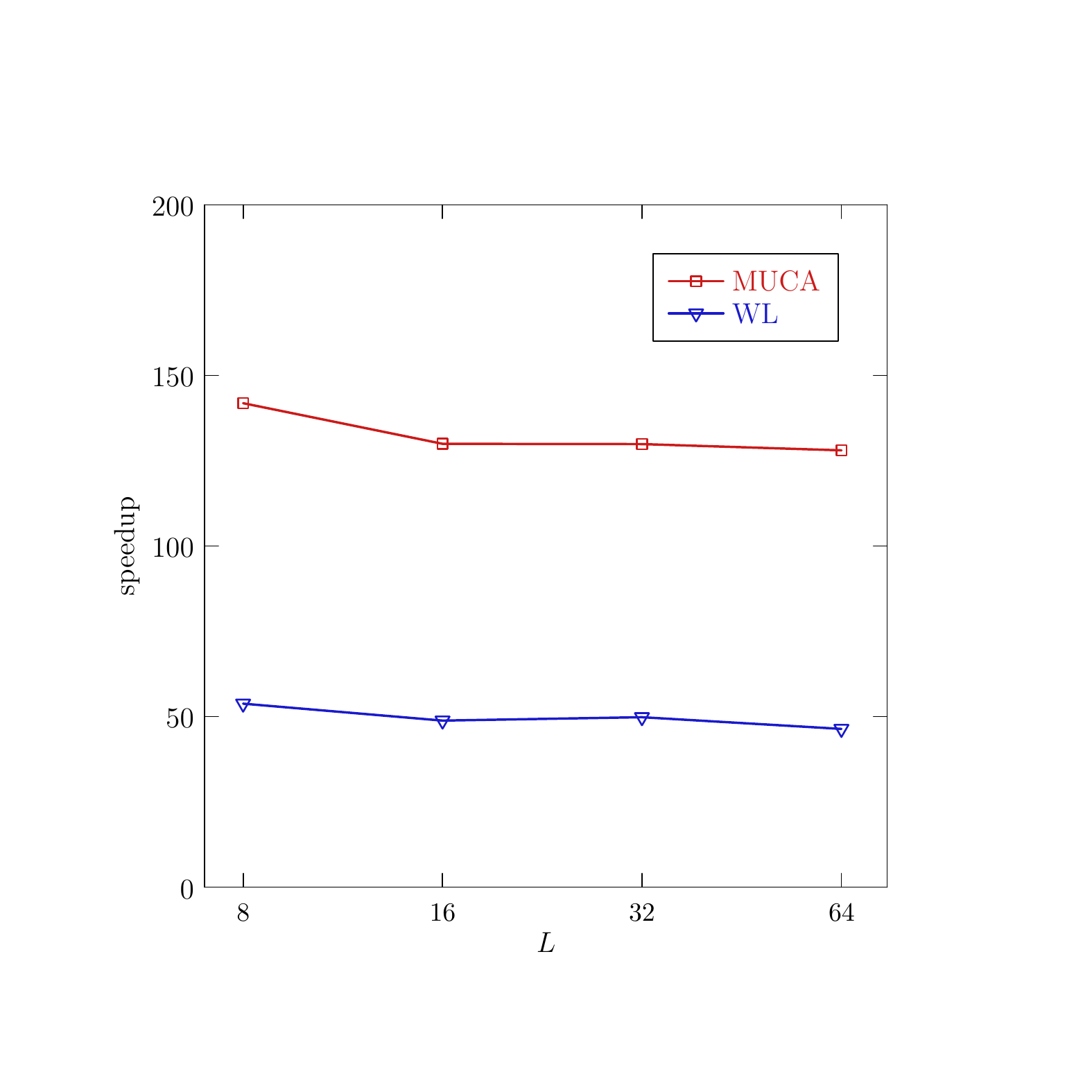}
  \caption{Left: speed-up of a Swendsen-Wang cluster update simulation of the 2D
    Ising model on a GTX 480 GPU as compared to the reference implementation on an
    Intel Q9650 CPU. Right: speed-up of the windowed multicanonical (MUCA) and
    Wang-Landau (WL) simulations of the 2D Ising model on the GTX 480.}
  \label{fig:cluster}
\end{figure}

\section{Cluster updates}

While single spin-flip simulations on a fixed lattice appear to be near optimal
problems for the parallel compute model of GPUs, highly non-local updates such as the
cluster algorithms used to beat critical slowing down in ferromagnetic models close
to a continuous phase transition are significantly harder to efficiently implement in
parallel. To test this, we considered different implementations of the Swendsen-Wang
cluster algorithm \cite{swendsen-wang:87a} for the Ising model. An update consists of
the following steps:
\begin{enumerate}
\item Activate bonds between like spins with probability $p = 1-e^{-2\beta J}$.
\item Construct (Swendsen-Wang) spin clusters from domains connected by active
  bonds.
\item Flip independent clusters with probability $1/2$.
\end{enumerate}
While steps $1$ and $3$ are completely local and hence can be easily performed in a
highly parallel fashion using a single thread for updating a few bonds or sites, the
cluster identification step is intrinsically non-local, in particular close to the
critical point where the Swendsen-Wang clusters undergo a percolation transition. The
approach we adopt uses two steps: (a) identify clusters in tiles of the system,
disregarding any couplings crossing the tile boundaries and (b) amalgamate those
small clusters by successively taking the boundary bonds into account. From the
approaches for clustering in tiles we implemented, including the Hoshen and Kopelman
algorithm, breadth-first search, union-and-find algorithms \cite{newman:01a} and
``self-labeling'' \cite{baillie:91}, we find the latter to be most efficient for
intermediate tile sizes. For self-labeling each site starts out with a unique cluster
label. Subsequently, each site in parallel sets its cluster label to the minimum of
its own label and that of its northward and eastward neighbors (in two
dimensions). This procedure is iterated until no change in cluster labels occurs. A
maximum of the order of $T$ iterations is required for tiles of size $T$, such that
in total $T^3$ operations are necessary. This appears unfavorable compared to the
other algorithms for local cluster labeling, but this disadvantage is
counter-balanced for tile sizes up to about $T=16$ by the high degree of
parallelism. From a number of different options for the consolidation of cluster
labels between tiles, including an iterative relaxation procedure and a hierarchical
approach based on union-and-find data structures, we find the latter to be
significantly more efficient. As is seen from the left panel of
Fig.~\ref{fig:cluster}, this allows for up to around 20-fold speed-ups as compared to
the CPU reference code, which appears respectable in view of the lack of locality,
but is clearly less impressive than the results for local simulations.

\section{Multicanonical and Wang-Landau simulations}

Simulations of systems undergoing first-order phase transitions or featuring complex
free-energy landscapes suggest the use of generalized-ensemble methods such as
multicanonical (MUCA) \cite{berg:92} or Wang-Landau (WL) \cite{wl:01}
simulations. Considering the internal energy $E$ as reaction coordinate, the
canonical distribution $p_\beta(E) = Z_\beta ^{-1} \Omega(E) \exp(-\beta E)$ is
generalized to read $p_\beta(E) = Z_\beta ^{-1} \Omega(E)/W(E)$. A flat histogram is
reached if the weights $W(E)$ equal the density of states $\Omega(E)$ or,
equivalently, $\omega(E) \equiv \ln W(E) = S(E)$, where $S(E)$ is the microcanonical
entropy. While MUCA uses a series of fixed-ensemble, equilibrium simulations to
estimate $W(E) = S(E)$, an analogous estimate is calculated online in a
non-equilibrium simulation in the WL approach. These algorithms are difficult to
parallelize since they require knowledge of the current value of a global reaction
coordinate (such as energy or magnetization) prior to each update. This effectively
serializes all updates performed on a single instance of the system. To still benefit
from the parallel GPU architecture, we use ``windowing'', i.e., the idea of applying
algorithms separately in small, fixed energy windows \cite{wl:01} and gluing together
the resulting estimates to reconstruct the overall $S(E)$.  We also use trivial
parallelization to improve statistics and estimate statistical errors.

We implemented MUCA and WL codes on GPU, where each thread works on a separate copy
of the system in a fixed individual energy window. Adjacent windows overlap by one
energy value. To construct an initial spin configuration inside of the desired window
we randomly select one among two nearest (AF or FM) ground states, and change its
energy by the steepest ascent method for randomly selected spins. Each block of
threads works on the same energy window to minimize thread divergence. Energy
histograms and $S(E)$ estimates are stored in shared memory. Spin configurations are
either directly worked on in global memory or tiled in shared memory. For a
sufficiently high load of the GPU and ensuring coalescence of memory accesses, we do
not find any benefit of using shared memory as each spin is only updated once before
it is written back to global memory. All calculations are done in single precision,
leading to essentially identical results as the double-precision CPU implementation.

For the 2D Ising model, we find that ``windowing'' does not cause systematic
deviations from the exact result for $S(E)$ \cite{beale:96} as long as enough
statistics is collected in each window. For the WL algorithm this means imposing a
strict criterion as to when the energy histogram is considered flat; for the MUCA
algorithm a sufficient number of tunneling events should be demanded. This allows us,
for instance, to construct $S(E)$ for a $64\times 64$ system from windows as small as
${\Delta E}=16$. The speedups of the GPU implementations using a sufficiently large
number of windows and independent runs to fully load the GPU are depicted in the
right panel of Fig.~\ref{fig:cluster} and summarized in Table
\ref{tab:performance}. For MUCA, we arrive at a speedup of $128$, similar to the
results found for the local algorithms, whereas the WL approach, in its current
implementation, allows a $46$ times performance increase only. This difference results
from the dynamical nature of the WL algorithm, where run times are random variables,
which leads to thread divergence and idle cores on the GPU. A more sophisticated
implementation using some load balancing scheme will be discussed elsewhere.

\begin{table}
  \centering
  \begin{tabular}{lllrrrr} \hline
    &                 &      & CPU         & C1060       & GTX 480&         \\
    System        & Algorithm       & $L$  & ns/flip     & ns/flip     & ns/flip&speed-up \\ \hline
    2D Ising      & Metropolis      & 32   &   8.3       &  2.58       & 1.60   &3/5      \\
    2D Ising      & Metropolis      & 16\,384 &  8.0     &  0.077      & 0.034  &103/235  \\
    3D Ising      & Metropolis      & 512  &  14.0       &  0.13       & 0.067  &107/209     \\
    2D Heisenberg & Metro.\ double  & 4096 & 183.7       &  4.66       & 1.94   &39/95      \\
    2D Heisenberg & Metro.\ single  & 4096 & 183.2       &  0.74       & 0.50   &248/366    \\
    2D Heisenberg & Metro.\ fast math   & 4096 & 183.2   &  0.30       & 0.18   &611/1018   \\ \hline
    2D spin glass & Metropolis      & 32   &  14.6       &  0.15       & 0.070   &97/209     \\
    2D spin glass & Metro. multi-spin  & 32  &  0.18     &  0.0075     & 0.0023  &24/78      \\ \hline
    2D Ising      & Cluster         & 8192 & 77.4        &   ---       & 4.62   &--/17  \\ \hline
    2D Ising      & multicanonical  & 64   & 42.1        &   ---       & 0.33  & --/128  \\ \hline
    2D Ising      & Wang-Landau     & 64   & 43.6        &   ---       & 0.94  & --/46  \\ \hline    
  \end{tabular}
  \caption{Spin-flip times for simulations of various lattice spin models with
    different algorithms on an Intel Q9650, a Tesla C1060 and a GTX 480,
    respectively. Apart from the cluster, multicanonical and Wang-Landau simulations,
  multi-hit updates with $k=100$ have been employed.}
  \label{tab:performance}
\end{table}

\section*{Acknowledgments}

The authors acknowledge support by the ``Center for Computational Sciences in Mainz''
(SRFN). M.W.\ received support by the DFG through the Emmy Noether
Programme under contract No.\ WE4425/1-1.



%\bibliographystyle{elsarticle-num}
%\bibliography{citeulike}

\end{document}